\documentclass[aps,groupedaddress,twocolumn]{revtex4}
\usepackage{amsmath,amsfonts}

\renewcommand{\vec}[1]{\underline{#1}}
\newcommand{\mat}[1]{\mathbf{#1}} 

\newcommand{\group}[1]{\mathbf{#1}}
\newcommand{\functional}[1]{\mathcal{#1}}
\newcommand{\operator}[1]{\hat{\mathrm{#1}}}

\newcommand{\Ket}[1]{|#1\rangle}

\newcommand{\Bracket}[3][ ]{\langle #2 | #1  #3\rangle}

\newcommand{\ignore}[1]{} 

\begin{document}


\title{Mapping the Generator Coordinate Method to the Coupled Cluster Approach}

\author{Jason L. Stuber} 
\email{bellygrub@phds.ca}
\noaffiliation


\date{2015.06.23}

\begin{abstract}
The generator coordinate method~(GCM) casts the wavefunction 
as an integral over a weighted set of non-orthogonal 
single determinantal states. In principle this representation
can be used like the configuration interaction~(CI) or shell model
to systematically improve the approximate wavefunction towards 
an exact solution. In practice applications have 
generally been limited to systems with less than three degrees of freedom. 
This bottleneck is directly linked to 
the exponential computational expense associated with the numerical projection of  
broken symmetry Hartree-Fock~(HF) or Hartree-Fock-Bogoliubov~(HFB) 
wavefunctions and to the use of a variational rather than a bi-variational
expression for the energy. We circumvent these issues by choosing a hole-particle 
representation for the generator and applying algebraic symmetry 
projection, via the use of tensor operators and the invariant mean (operator average). 
The resulting GCM formulation can be mapped directly to the coupled cluster~(CC) 
approach, leading to a significantly more efficient approach than the 
conventional GCM route. 
\end{abstract}


\maketitle



\section{The Generator Coordinate Method~(GCM)}
The generator coordinate method~(GCM),  
initially developed to treat large amplitude collective motion in 
nuclei~\cite{hillwheeler:1953,griffinwheeler:1957,griffin:1957} (see also Refs.~\cite{reinhardgoeke:1987,ringschuck:1980}), represents 
the wavefunction as an integral over a weighted set of states
\begin{align}
\Ket{\Psi_\mathrm{GCM}}&= \int d\mat{Z} \,  f(\mat{Z})  \, \Ket{\Phi(\mat{Z})} \, ,
\label{eq:GCM}
\end{align}
each state $\Ket{\Phi(\mat{Z})}$ parameterized by a continuous variable $\mat{Z}$, 
the so-called generator coordinate. For bosonic modes this structural form is
 related to the coherent state representation~\cite{glauber:1963}. 
Unlike the configuration interaction~(CI) or shell model
\begin{align}
\Ket{\Psi_{\mathrm{CI}}} & = \sum_{\mu} \, C_{\mu} \, \Ket{\Phi_{\mu}} \, ,
\label{eq:CI}
\end{align}
which expands the wavefunction as a linear combination of mutually orthogonal states
\begin{align}
\Bracket{\Phi_{\mu}}{\Phi_{\nu}} = \delta_{\mu\nu} \, ,
\end{align} 
the states $\Ket{\Phi(\mat{Z})}$ which comprise the GCM are not mutually orthogonal
by design, namely the state overlap matrix 
\begin{align}
\mat{N}(\mat{Z}';\mat{Z}) &= \Bracket{\Phi(\mat{Z}')}{\Phi(\mat{Z})} \, ,
\end{align}
is not diagonal. 

Of primary interest in this work are fermions (electrons or nucleons)
for which each basis state $\Ket{\Phi(\mat{Z})}$ is then 
a Slater determinant, weighted by the complex valued scalar function $f(\mat{Z})$
in  Eq.~\eqref{eq:GCM}.
Thouless' 
Theorem~\cite{thouless:1960,thouless:1961} parametrizes these determinants 
by an exponential acting on a reference state $\Ket{\Phi_{0}}$ 
\begin{align}
\Ket{\Phi(\mat{Z})} &= e^{\operator{Z}} \, \Ket{\Phi_{0}} \, , 
\end{align}
where, schematically 
\begin{align}
\operator{Z} & = \sum_{\mu} Z_{\mu} \operator{E}_{\mu} \, , \label{eq:Zmu}
\end{align}
is a mono-excitation level generator 
with corresponding coefficients $\mat{Z}=(Z_{\mu})$,
$\operator{E}_{\mu}$ is the product of two fermion
creation and/or annihilation operators.
The structure of $\operator{Z}$ and $\Ket{\Phi_{0}}$ in turn
define the space spanned by all possible GCM wavefunctions. 
If the generators $\operator{Z}$ are not sufficiently general, 
breaking one or more symmetries of the hamiltonian, then 
it is not possible to represent all symmetry preserving wavefunctions. 

Solutions for the GCM wavefunction are obtained in the time-independent
case via the Schr\"{o}dinger equation  
\begin{align}
 \operator{H} \, \Ket{\Psi_{\mathrm{GCM}}} & = E \, \Ket{\Psi_{\mathrm{GCM}}} \, ,
\end{align}  
which can be solved via a matrix representation, resulting in the 
Hill-Wheeler-Griffin~(HWG) equation
\begin{align}
 \int d\mat{Z} \,   \mat{H}(\mat{Z}';\mat{Z}) \, f(\mat{Z})
  & = E \int d\mat{Z} \, \mat{N}(\mat{Z}';\mat{Z}) \, f(\mat{Z})  \, ,
  \label{eq:continuousSE} 
\end{align}
which is a generalized eigenvalue problem in continuous index form.
Here
\begin{align}
 \mat{H}(\mat{Z}';\mat{Z}) &= \Bracket[\, \operator{H} \, |]{\Phi(\mat{Z}')}{\Phi(\mat{Z})} 
 \label{eq:GCM_hamiltonian} \\
\mat{N}(\mat{Z}';\mat{Z}) &= \Bracket{\Phi(\mat{Z}')}{\Phi(\mat{Z})} \, .\label{eq:GCM_overlap}
\end{align}
A strength of the GCM representation is that evaluating the matrix 
elements Eq.~\eqref{eq:GCM_hamiltonian},~\eqref{eq:GCM_overlap},  
is inexpensive, even between single Slater determinants corresponding to
different basis sets or when also evaluating operators between such states, 
by using the L\"{o}wdin or Onishi 
formulae~(see Ref.~\cite{ringschuck:1980} for the combined expression and original references).

In practice, the generalized eigenvalue problem Eq.~\eqref{eq:continuousSE} 
can be solved by explicit discretization. 
In this case the matrix formulation is given by
\begin{align}
\mat{H} \,\vec{f} & = E \,\mat{N} \,\vec{f} \, , \label{eq:generalized_eigenvalue} 
\end{align}
for $K$ sample points $\mat{Z}^{(1)},\ldots,\mat{Z}^{(K)}$ and the, to be determined, weightings 
$f_k= f(\mat{Z}^{(k)})$. Using a naive discretization, the 
computational demand scales
exponentially with the dimension $D$, the number of 
generator coordinates $\mat{Z}$.
If one takes modestly $N\approx 10$ points per dimension then the 
number of samples is $K=N^D = 10^D$, quickly becoming intractable.

Applications in nuclear structure tend to use only a few  
( $ D \leq 3 $ ) generator coordinates, 
selecting only those operators (such as the quadrupole operator) 
which correspond to physically motivated nuclear deformation modes in the 
expansion, rather than the full space of mono-excitation operators. 
This initial choice was due to the phenomenological origin of the GCM
and applications of interest. Subsequent investigations 
over the last sixty years have been hampered 
in no small part due to the poor scaling behaviour with the 
increase in the number of dimensions. This is also true even
when the GCM formalism is only used as a means of symmetry 
projection for even a single broken symmetry Hartree-Fock~(HF) 
or Hartree-Fock-Bogoliubov~(HFB) state and the weightings 
$f(\mat{Z})$ are fixed by symmetry. Other approaches have
parameterized the weighting function $f(\mat{Z})$ by 
a combination of gaussian functions to restore some of the 
continuous flavour of the GCM ansatz, however the underlying
numerical issues still remain.

To circumvent these difficulties, we  
propose an alternate formulation of the GCM. This
is possible by using a hole-particle representation
of the generator coordinate operators which then allows for
analytic symmetry projection using tensor operators
and the invariant mean (operator average)~\cite{naimark:1982,stuber:2009}.
Symmetry projection
then becomes a ``back of an envelope calculation''. 
Explicitly, such mono-excitation 
generators $\operator{Z}$ of the hole-particle type are
defined as 
\begin{align}
\operator{Z} &= \sum_{AI} Z_{AI} \, \operator{a}_{A}^{\dag} \operator{a}_{I} \, ,
\label{eq:HFgenerator}
\end{align}
when generating the Hartree-Fock~(HF) class of states $\Ket{\Phi(\mat{Z})}$ and 
\begin{align}
\operator{Z} &= \sum_{AI} \, Z_{AI} \, \operator{a}_{A}^{\dag}\operator{a}_{I} \nonumber \\
 & + \tfrac{1}{2}\sum_{AB} \, Z_{AB}^{*} \, \operator{a}_{A}^{\dag}\operator{a}_{B}^{\dag}
   + \tfrac{1}{2}\sum_{IJ} \, Z_{IJ} \, \operator{a}_{I}\operator{a}_{J} \, ,
\label{eq:HFBgenerator}
\end{align}
when generating the Hartree-Fock-Bogoliubov~(HFB) class of states. 
Here $I$, $J$ index the occupied spin orbitals (holes) and $A$, $B$ the 
unoccupied spin orbitals (particles)  
with respect to the single-determinantal reference $\Ket{\Phi_{0}}$. 

The resulting GCM formulation has a direct connection with 
projected Hartree-Fock~(PHF) and projected 
Hartree-Fock-Bogoliubov~(PHFB) wavefunctions, which  
then correspond to the single term approximation of 
the explicitly projected GCM. 
It is also related to the coupled cluster~(CC) 
approach~\cite{cizek:1966,cizek:1969} (see also Refs.~\cite{paldus:1992,paldus:1999}). 
A mapping from
the general GCM prescription to CC can be constructed, even when 
no explicit symmetry breaking is present. This mapping between 
GCM and CC leads to a distribution-like parameterization of the 
cluster amplitudes, and an alternate route to solve for the GCM
wavefunction via the CC equations. The following sections will develop 
this new prescription.

\section{Mapping From GCM to CC}
In this section a mapping from the GCM representation 
\begin{align}
\Ket{\Psi_\mathrm{GCM}}&= \int d\mat{Z} \,  f(\mat{Z})  \, e^{\operator{Z}} \, \Ket{\Phi_{0}} \, ,
\label{eq:GCM_wavefunction}
\end{align}
to the coupled cluster representation 
\begin{align}
\Ket{\Psi_{\mathrm{CC}}} & = e^{\operator{T}} \, \Ket{\Phi_{0}} \, , 
\label{eq:CC_wavefunction}
\end{align}
of the wavefunction will be elucidated. Here
\begin{align}
\operator{T} & = \operator{T}_{1} + \operator{T}_{2} + \operator{T}_{3}+\operator{T}_{4}
+ \ldots   \label{eq:cc_expansion}
\end{align}
are the conventional mono-, bi- and higher order cluster excitation 
operators~(see {\em e.g.} Refs.~\cite{paldus:1992,paldus:1999}).
Explicitly,
\begin{align}
\exp(\operator{T}) & = 1  \nonumber \\
& + \operator{T}_{1} \nonumber \\
& +\operator{T}_{2}+\frac{1}{2}\operator{T}_{1}^{2}  \label{eq:explicitCC}\\
 &+ \operator{T}_{3}+\operator{T}_{2}\operator{T}_{1}+\frac{1}{3!}\operator{T}_{1}^{3}
 \nonumber \\
 & 
+\operator{T}_{4} +\operator{T}_{3}\operator{T}_{1}
+\frac{1}{2!}\operator{T}_{2}^2 + \frac{1}{2!}\operator{T}_{2}\operator{T}_{1}^2
+\frac{1}{4!}\operator{T}_{1}^{4} +\ldots \nonumber \, ,
\end{align}
where terms corresponding to the same excitation order have been grouped.
The mapping from GCM to CC can be expressed from two different perspectives
on how the symmetry projection is accomplished: either
when the weightings $f(\mat{Z})$ are constrained via external means
(such as via Lagrange multipliers), or when the weightings 
are unconstrained but an explicit projection has been performed,
which the authors recommended as it is the most computationally efficient. 
These mappings from GCM to CC 
are also possible even when symmetry projection is not present or required.

\subsection{Constrained Weighting Formulation}
The GCM wavefunction Eq.~\eqref{eq:GCM_wavefunction} is generally
formulated in terms of broken symmetry generators $\operator{Z}$ 
as in Eqs.~\eqref{eq:HFgenerator} or \eqref{eq:HFBgenerator}
however the final wavefunction $\Ket{\Psi_{\mathrm{GCM}}}$ is 
also implicitly assumed to have good quantum numbers and therefore
be symmetry adapted.
Unless an explicit projection has already been performed (as will be 
illustrated in the next section), then the weightings $f(\mat{Z})$ must
satisfy additional constraints to ensure symmetry is preserved for
the overall wavefunction. This can be accomplished by  
an implicit parameterization (if possible) 
or by the use of additional constraints imposed 
via Lagrange multipliers. In either approach, the resulting GCM wavefunction
is then given by
\begin{align}
\Ket{\Psi_\mathrm{GCM}}&= \int d\mat{Z} \,  f_{c}(\mat{Z})  \, e^{\operator{Z}}  \,\Ket{\Phi_{0}} \, ,
\label{eq:constrainedGCM}
\end{align}
where $f_c(\mat{Z})$ now denote the constrained weights 
which ensure symmetry is preserved. Possible methods to determine
these weights include by explicit incorporation of the projection via 
a Baker-Campbell-Hausdorff~(BCH) expansion (which also 
effectively modifies $\operator{Z}$)
or
by requiring the symmetry violating contributions vanish at each order; similar to 
the Brillouin conditions for HFB. How these weights can be obtained will not 
however be considered further in this work. 
Instead a mapping to the coupled cluster form of the wavefunction will now be developed. 

The constrained weighting formulation of GCM Eq.~\eqref{eq:constrainedGCM} 
can be compared with the coupled cluster representation Eq.~\eqref{eq:CC_wavefunction}, 
\begin{align}
e^{\operator{T}} \, \Ket{\Phi_{0}} & = \int d\mat{Z} \,  f_{c}(\mat{Z})  \, e^{\operator{Z}} \, \Ket{\Phi_{0}} \, , 
\end{align} 
for each excitation order $\mat{\operator{Z}}^k$, as compared to
the corresponding $k$-th row in the cluster expression in Eq.~\eqref{eq:explicitCC}. 
At zeroth order,
\begin{align}
1 & = \int d\mat{Z} \,  f_{c}(\mat{Z}) \, , \label{eq:ZerothOrder}
\end{align}
one obtains a normalization condition on the constrained weights $f_{c}(\mat{Z})$. 
One is tempted to treat $f_{c}(\mat{Z})$ as a distribution, however the scalar weights 
are themselves generally complex-valued and even when real valued can 
have regions in the parameter space where the weight is negative.

Turning to the first order term in the expansion,
\begin{align}
\operator{T}_{1}  & = \int d\mat{Z} \, f_c(\mat{Z}) \,  \operator{Z} \, ,  \label{eq:t1} 
\end{align}
this equality suggests the interpretation that under the GCM to CC mapping, the  
coupled cluster mono-excitation operator $\operator{T}_{1}$ corresponds to an average 
over the `pseudo-density' $f_{c}(\mat{Z})$. Explicitly, in terms of the generator coordinate parameters, with $\operator{T}_{1}$ defined in a similar fashion
 to Eq.~\eqref{eq:Zmu},
\begin{align}
T_{\mu} & = \int dZ_{1}\ldots dZ_{\mu}\ldots dZ_{D} \, f_c(\mat{Z}) \, Z_{\mu} \nonumber\\
 & = \int dZ_{\mu} \, f_{c}^{(\mu)}(Z_{\mu}) \, Z_{\mu} \, ,
\end{align}
where
\begin{align}
f_{c}^{(\mu)}(Z_{\mu}) 
& = \int dZ_{1}\ldots dZ_{\mu -1}dZ_{\mu +1}\ldots dZ_{D} \, f_c(\mat{Z})  \, ,
\end{align}
is a reduced (constrained) weight obtained by 
integrating over all generator coordinates $\mat{Z}$ 
except for $Z_{\mu}$. From the normalization condition 
Eq.~\eqref{eq:ZerothOrder}, then
\begin{align}
\int dZ_{\mu} \, f_{c}^{(\mu)}(Z_{\mu})  & = 1 \, ,
\end{align}
for all choices of mono-excitation index $\mu$. This again suggests 
the interpretation that $f_{c}(\mat{Z})$ can be considered a
(complex valued) distribution, in 
which case $f_{c}^{(\mu)}(Z_{\mu})$ is then a one-parameter reduced 
weight.

At quadratic ($\operator{Z}^{2}$) order in the expansion,
\begin{align}
\left( \operator{T}_{2}+\frac{1}{2} \operator{T}_{1}^{2} \right)  & =
\frac{1}{2}\int d\mat{Z} \, f_c(\mat{Z}) \,  \operator{Z}^{2}  \, . 
\end{align}
Rewriting
\begin{align}
\operator{T}_{2} & = 
\frac{1}{2}\int d\mat{Z} \, f_c(\mat{Z}) \, \operator{Z}^2 - \frac{1}{2}\operator{T}_{1}^{2}
\nonumber \\
&=\frac{1}{2}\int d\mat{Z} \, f_c(\mat{Z}) \, (\operator{Z}-\operator{T}_{1})^2 \, , 
\end{align}
The bi-excitation operator $\operator{T}_{2}$ obtained from the 
constrained weighting GCM to CC mapping
is the fluctuation (deviation) from the average. 
As before, one can formulate a two-parameter reduced 
weighting $f^{(\mu\nu)}(Z_{\mu},Z_{\nu})$ for $\mu\neq\nu$. 

Continuing in a similar fashion, one finds  the higher order excitations 
\begin{align}
\operator{T}_{k} & = \frac{1}{k!}\int d\mat{Z} \, f_c(\mat{Z}) \, 
(\operator{Z}- \operator{T}_{1})^k \, , 
\end{align}
as the moments (or cumulants) in the space of mono-excitations 
about the mean given by $\operator{T}_{1}$. 
This is the underlying structure of GCM, as cast in the constrained 
weighting formulation. 

\subsection{Explicit Projection Formulation}
One difficulty with the approach taken in the previous section 
is that the constrained weights
$f_{c}(\mat{Z})$ require either a specific parametrization or 
the use of external constraints, 
increasing the difficulty for
a practical solution. This constraint 
on the weightings can be removed if explicit 
projection is used. The weights $f(\mat{Z})$ are then 
unconstrained and generally will not have the same values
as $f_c(\mat{Z})$, although a mapping between these quantities 
is possible. 

Consider the explicit projection,
\begin{align}
\operator{P} \, \Ket{\Psi_{\mathrm{GCM}}} & = \operator{P} \, 
\int d\mat{Z} \, f(\mat{Z}) \,  e^{\operator{Z}} \, \Ket{\Phi_{0}} \nonumber \\
 & = 
\int d\mat{Z} \, \left( \operator{P} \, f(\mat{Z}) \,  e^{\operator{Z}} \, \Ket{\Phi_{0}}
\right) \, .
\end{align}
If the GCM wavefunction is already symmetry adapted, such as by using constrained
weights $f_{c}(\mat{Z})$, then this is identical to the original 
expression Eq.~\eqref{eq:GCM_wavefunction}. If this is not the
case, then evaluation of the
projection even for a single state,  
\begin{align}
\operator{P} \, \Ket{\Phi(\mat{Z})} 
& = \operator{P} \,   e^{\operator{Z}} \, \Ket{\Phi_{0}} \, ,
\end{align}
can be extremely computationally demanding or even intractable,
depending on the symmetry group $\group{G}$ of the projection 
operator $\operator{P}$ and which subgroup $\group{H}\subset\group{G}$ 
the generators $\operator{Z}$ transform 
under~\cite{fukutome:1982,stuber:2002,stuber:2003}. 
Spin angular momentum $\group{G}=\group{SU}(2)$ is the
first example of such a `problem group', where exact 
state projection methods scale with the size of the state space, i.e.
the number of possible Slater determinantal configurations.

If the reference $\Ket{\Phi_{0}}$ is however already symmetry adapted 
and $\operator{Z}$ is of the hole-particle type (as we have explicitly chosen),
then one can determine an analytical form for the symmetry projection
through the use of tensor operators and the invariant 
mean (operator average)~\cite{naimark:1982,stuber:2009}. 
Instead of the state projection or approximate numerical projection, 
one employs operator projection
\begin{align}
\operator{U}_{\mathrm{sa}} & = \functional{M}_{\group{G}}(e^{\operator{Z}}) \, , 
\label{eq:OperatorProjection}
\end{align}
in which case
\begin{align}
\operator{P} \,   e^{\operator{Z}} \, \Ket{\Phi_{0}}
 & = \operator{U}_{\mathrm{sa}} \Ket{\Phi_{0}} \, . \label{eq:invariant_assumption}
\end{align}
Here $\functional{M}_{\group{G}}(\cdot)$ is the invariant 
mean, which for discrete groups $\group{G}$ corresponds 
to the average
\begin{align}
\functional{M}_{\group{G}}(\operator{A})& = \frac{1}{|\group{G}|} \sum_{\operator{g}\in\group{G}}
 \operator{g} \, \operator{A} \, \operator{g}^{-1} \, . 
\end{align}
For $\group{SU}(2)$ spin projection, see Ref.~\cite{stuber:2009}.
This equality Eq.~\eqref{eq:invariant_assumption} only
holds under certain assumptions on the quantum numbers,
for the case of $\group{SU}(2)$ that the reference is a 
closed-shell singlet $S=0$, however this method can be extended
to non-singlet quantum numbers simply by inclusion of additional 
tensor operator coupling terms weighted with the appropriate
Clebsch-Gordon coefficient. The operator projection 
Eq.~\eqref{eq:OperatorProjection} then can be expanded for each 
order in $\operator{Z}$, and significantly for each order the symmetry projection
is algebraic and exact. 

The structure of the symmetry adapted correlation operator $\operator{U}_{\mathrm{sa}}$
can be formally linked to the coupled cluster expansion,
so that
\begin{align}
\operator{U}_{\mathrm{sa}} & =\exp(\operator{S}) \, ,  \label{eq:US}
\end{align}
where
\begin{align}
\operator{S}& = \operator{S}_{1} + \operator{S}_{2} + \operator{S}_{3} + \operator{S}_{4} + \ldots \label{eq:Sexpansion}
\end{align}
are the symmetry adapted excitations, in correspondence with Eq.~\eqref{eq:cc_expansion}.
Comparing Eqs.~\eqref{eq:OperatorProjection} with \eqref{eq:US}, then
\begin{align}
\operator{S}_{1} & = \functional{M}_{\group{G}}(\operator{Z}) \label{eq:s1}\\
\operator{S}_{2} & = \functional{M}_{\group{G}}\left(\frac{1}{2}\operator{Z}^{2}\right) 
 - \frac{1}{2}\operator{S}_{1}^{2} \\
\operator{S}_{3} & = \functional{M}_{\group{G}}\left(\frac{1}{3!}\operator{Z}^{3}\right) 
 -\operator{S}_{2}\operator{S}_{1}- \frac{1}{3!}\operator{S}_{1}^{3} \\
\operator{S}_{4} & = \functional{M}_{\group{G}}\left(\frac{1}{4!}\operator{Z}^{4}\right) 
 -\operator{S}_{3}\operator{S}_{1}
 -\frac{1}{2}\operator{S}_{2}^{2} 
 -\frac{1}{2}\operator{S}_{2}\operator{S}_{1}^{2}- \frac{1}{4!}\operator{S}_{1}^{4}  \, , 
 \label{eq:s4}
\end{align}
and so on.  This approach was advocated in our 
previous work~\cite{stuber:2009} as an exact method at any order 
to analytically determine projected broken symmetry wavefunctions,
including for higher order operators approximations ({\em e.g.} PUCCSD). 
Once the symmetry group $\group{G}$ 
is known, explicit forms for the invariant 
mean $\functional{M}_{\group{G}}$ are relatively straightforward to evaluate
for any $k$-body operator, particularly when $\operator{Z}$ 
is represented in terms of tensor operators.

As a simple example of this symmetry projection, 
consider the generator $\operator{Z}$ corresponding to real-valued BCS states,
\begin{align}
\operator{Z}&= \sum_{N,S,M} \operator{Z}^{(N,S,M)}
\nonumber \\
& = \sum_{N=-1}^{+1} \operator{Z}_{R}^{(N,0,0)} \, ,
\end{align}
given in terms of its tensor components $\operator{Z}^{(N,S,M)}$. 
The corresponding cluster operators $\operator{S}$ are given by
\begin{align}
\operator{S}_{1} & = \operator{Z}_{R}^{(0,0,0)} \\
\operator{S}_{2} & = \operator{Z}^{(1,0,0)}_{R} \, \operator{Z}^{(-1,0,0)}_{R} \\
\operator{S}_{3} & = 0 \\
\operator{S}_{4} & = -\frac{1}{4} 
\left(\operator{Z}_{R}^{(1,0,0)}\right)^2 \left(\operator{Z}_{R}^{(-1,0,0)} \right)^2 
\nonumber \\
 & = -\frac{1}{4} \, \operator{S}_{2}^2 \, ,
\end{align}
which can be verified via Eqs.~\eqref{eq:s1}--\eqref{eq:s4}.
This can similarly be applied to other symmetry groups. Explicit expressions for 
$\functional{M}_{\group{G}}(\cdot)$ operator projection of up to fourth order will be 
provided in a forthcoming paper~\cite{stuber:2015b} for the symmetry group 
$\group{G}=\group{S}\times\group{T}\times\group{N}$, which corresponds to the 
projected real and complex-valued RHF, UHF, GHF, BCS and HFB classes of 
wavefunctions.

For the remainder of this paper it is taken that the assumptions required
for the invariant mean projection Eq.~\eqref{eq:invariant_assumption} to be
valid hold. For specificity, in case of the quantum chemistry literature 
this would require that the orbitals are real valued canonical spin-orbitals 
(space $\times$ spin)
and that the reference $\Ket{\Phi_{0}}$ is a closed shell singlet state 
with good particle number. 
For simplicity, the generator coordinates $\operator{Z}$ and weightings 
$f(\mat{Z})$ will also be assumed  real valued. These conditions are 
intended to simplify the derivation of this alternate GCM prescription, 
avoiding the potential quagmire with spin coupling and other issues for which a unique 
definition of the hole-particle representation is not as straightforward. 
Some of these conditions can be relaxed and the GCM structural form 
generalized further, such as by allowing for different reference states 
or by choosing $\Ket{\Phi(\mat{Z})}$ to include both single and 
double excitations for the generator coordinates. 
The authors however wish to first grab the proverbial low-lying fruit first.

Using the operator average, then 
\begin{align}
\operator{P}\Ket{\Psi_{\mathrm{GCM}}} & = \operator{P}
\int d\mat{Z} \, f(\mat{Z}) \,  \functional{M}_{\group{G}} (e^{\operator{Z}}) \, \Ket{\Phi_{0}}
\, ,  
\end{align}
or using Eqs.~\eqref{eq:invariant_assumption}, \eqref{eq:US},
\begin{align}
\operator{P}\Ket{\Psi_{\mathrm{GCM}}}
&= \int d\mat{Z} \, f(\mat{Z}) \, e^{\operator{S}(\operator{Z})}\, \Ket{\Phi_{0}}  \, .\label{eq:GCMasPHF}
\end{align} 
In this representation, the GCM is a 
(continuous) linear combination of cluster operators, each 
corresponding to a projected broken symmetry  wavefunction. 

As before, comparing order by order with the coupled cluster expansion
\begin{align}
1 & = \int d\mat{Z} \, f(\mat{Z}) \, , 
\end{align}
but now for the unconstrained weightings. 
The mono-excitation operator is given by
\begin{align}
\operator{T}_{1} & = \int d\mat{Z} \, f(\mat{Z}) \, \functional{M}_{\group{G}}(\operator{Z})
\nonumber \\
 & = \int d\mat{Z} \, f(\mat{Z}) \, \operator{S}_{1}(\operator{Z})  \, , 
\end{align}
and so the unconstrained mono-excitation $\operator{T}_{1}$ is an average over
projected mono-excitation generator $\operator{Z}$. 

Turning to the bi-excitations, 
\begin{align}
\operator{T}_{2} +\frac{1}{2} \operator{T}_{1}^{2} & = \frac{1}{2}
\int d\mat{Z} \, f(\mat{Z}) \,
\functional{M}_{\group{G}}(\operator{Z}^{2})  \, ,
\end{align}
or
\begin{align}
\operator{T}_{2} & = \frac{1}{2!} \int d\mat{Z} \, f(\mat{Z}) \,
\functional{M}_{\group{G}}(\operator{Z}^{2})
 - \frac{1}{2!}\operator{T}_{1}^{2}  \, .
\end{align}
The bi-excitation can be interpreted, just as before, as the deviation
from the average amplitude
\begin{align}
\operator{T}_{2} & = \frac{1}{2!} \int d\mat{Z} \, f(\mat{Z}) \,
\functional{M}_{\group{G}}((\operator{Z}-\operator{T}_{1})^{2}) \, , \label{eq:GCMasPP}
\end{align}
where however the symmetry projection via the invariant mean $\functional{M}_{\group{G}}(\cdot)$ 
has been incorporated with unconstrained weightings $f(\mat{Z})$.

Alternately, the bi-excitations can be re-cast using the cluster expansion for each term,
\begin{align}
\operator{T}_{2} & = \int d\mat{Z} \, f(\mat{Z}) \,
\left( \frac{1}{2}\operator{S}_{1}^{2}(\operator{Z}) +\operator{S}_{2}(\operator{Z}) \right) 
-\frac{1}{2!} \operator{T}_{1}^{2} \nonumber \\
 & = \int d\mat{Z} \, f(\mat{Z}) \, \operator{S}_{2}(\operator{Z}) 
 \nonumber \\
 & + \frac{1}{2} \int d\mat{Z} \, f(\mat{Z}) \,
\left( \operator{S}_{1}(\operator{Z})-\operator{T}_{1}\right)^{2} \, ,
\label{eq:t2new}
\end{align}
in which case the $\operator{T}_{2}$ comes both 
from the sum of a weighted $\operator{S}_{2}$ 
and the deviation from the operator average of $\operator{T}_{1}$. Here the
$\operator{S}_{2}$ contribution comes from explicit
symmetry breaking, whereas the second term comes from the 
spread of the mono-excitation coefficients or alternately the
cross-correlation. The $\operator{S}_{2}$ 
term vanishes entirely if no symmetry breaking is present and 
can be interpreted as the canonical order parameter.
One finds that  
$\operator{T}_{2}$ only vanishes if both $\operator{S}_{2}=0$ and
\begin{align}
f(\mat{Z}) & = \delta(\mat{Z}-\mat{Z}_{0}) \, ,
\end{align}
that is that the GCM corresponds to a single symmetry restricted 
Slater determinant, {\em i.e.} restricted Hartree-Fock~(RHF)
wavefunction. 

At the $\operator{T}_{2}$ bi-excitation level,
GCM is a (continuous) linear combination of projected broken symmetry 
wavefunctions as in Eq.~\eqref{eq:GCMasPHF} or instead via Eq.~\eqref{eq:t2new} 
is a weighted distribution of pairing amplitudes. The latter 
scheme can be
related to the plethora of approximate doubles schemes, such as perfect pairing, 
RVB, GVB, canonical BCS, AGP, and so on. This representation 
thus provides a natural framework to discuss the
sparsity structure of the cluster amplitudes, which can be
dictated by symmetry or other considerations.

For the third order expression,
\begin{align}
\operator{T}_{3} + \operator{T}_{2}\operator{T}_{1}
+\frac{1}{3!}\operator{T}_{1}^{3} & = 
\frac{1}{3!} \int d\mat{Z} \, f(\mat{Z}) \functional{M}_{\group{G}}(\operator{Z}^{3}) \, .
\end{align}
Re-arranging, 
\begin{align}
\operator{T}_{3} & = \frac{1}{3!} \int d\mat{Z} \, f(\mat{Z})  \, 
 \functional{M}_{\group{G}}((\operator{Z}-\operator{T}_{1})^{3}) \, ,
\end{align}
which can be linked directly to the corresponding
symmetry constrained weighting $f_{c}(\mat{Z})$
expression. Alternately, in terms of the symmetry adapted cluster operators,
\begin{align}
\operator{T}_{3} & = 
 \int d\mat{Z} \, f(\mat{Z}) \, \operator{S}_{3}(\mat{Z}) \nonumber \\
 & + \int d\mat{Z} \, f(\mat{Z}) \, (\operator{S}_{2}(\mat{Z})-\operator{T}_{2})
 (\operator{S}_{1}(\mat{Z})-\operator{T}_{1}) \nonumber \\
  & + \frac{1}{3!}\int d\mat{Z} \, f(\mat{Z}) \, (\operator{S}_{1}(\mat{Z})-\operator{T}_{1})^{3} \, .
\end{align}
The expansion suggests that the higher order cluster amplitudes, when
interpreted through the GCM to CC mapping, are a combination of `pure' symmetry 
breaking contributions together with distributional (cross-correlation) contributions. 

Fourth order proceeds similarly,
\begin{align}
\operator{T}_{4} & = \frac{1}{4!} \int d\mat{Z} \, f(\mat{Z})  \, 
 \functional{M}_{\group{G}}((\operator{Z}-\operator{T}_{1})^{4}) \, ,
\end{align}
and for the cluster formulation
\begin{align}
\operator{T}_{4} & = 
\int d\mat{Z} \, f(\mat{Z}) \, \operator{S}_{4}(\mat{Z}) \nonumber \\
 & + \int d\mat{Z} \, f(\mat{Z}) \, (\operator{S}_{3}(\mat{Z})-\operator{T}_{3})
 (\operator{S}_{1}(\mat{Z})-\operator{T}_{1}) \nonumber \\
 &  + \frac{1}{2!}\int d\mat{Z} \, f(\mat{Z}) \, (\operator{S}_{2}(\mat{Z})-\operator{T}_{2})^{2} \nonumber \\
 &  + \frac{1}{2!}\int d\mat{Z} \, f(\mat{Z}) \, (\operator{S}_{2}(\mat{Z})-\operator{T}_{2}) \, (\operator{S}_{1}(\mat{Z})-\operator{T}_{1})^{2}  \nonumber \\
 &  + \frac{1}{4!}\int d\mat{Z} \, f(\mat{Z}) \, (\operator{S}_{1}(\mat{Z})-\operator{T}_{1})^{4}
\end{align}
This is the structure of the GCM under the explicitly projected CC mapping,
which provides an alternate route to a highly efficient and direct means 
of symmetry projection.

\section{First Checkpoint}
This work has provided a clear prescription for how to utilize the structural
form of GCM while avoiding the historical computational bottleneck.
This has been accomplished through the use of analytical projection
via the invariant mean and tensor operator representations, which can
be effected efficiently when the generator coordinate is cast within
the hole-particle representation. 

At the lowest level of approximation, which takes a single 
Slater determinant for the GCM, this approach yields an efficient 
method to obtain the wavefunction for individual PHF and PHFB states
as cast within the coupled cluster representation. The 
imposed sparsity on the cluster amplitudes (for
a single or linear combination of Slater determinants) can be formulated
in this framework,
and related to the canonical (finite size) order parameters of the system. 
For higher order amplitudes, the symmetry projection can also
be used to obtain the  externally corrected 
correlation corrections for the $\operator{T}_{3}$ and $\operator{T}_{4}$ 
amplitudes. These can then be determined self-consistently through the
higher order cluster equations, as significantly fewer 
free parameters are required as than a full $\operator{T}_{3}$ or $\operator{T}_{4}$
calculation.    

The GCM expansion provides a natural approach to go beyond projected HF and projected HFB 
wavefunctions towards the exact solution, and in this work has been mapped to 
the coupled cluster amplitudes. Under this mapping, the resulting 
GCM weights $f(\mat{Z})$ can be interpreted
as a pseudo-distribution. Provided the weights $f(\mat{Z})$ between
uncoupled subsystems are uncorrelated, then the GCM wavefunction 
is size-extensive, just as for the coupled cluster approach.

An explicit solution procedure has not been presented in this
work, however one can use the coupled cluster equations
directly as the authors have advocated previously for projected
broken symmetry wavefunctions~\cite{stuber:2009}. 
Even at the single term approximation of GCM, which 
corresponds to the PHF or PHFB wavefunctions, this leads to a 
reduction of the cluster equations to a relatively few expressions 
for the free parameters. This approach can also be used for GCM
even when symmetry 
breaking is not present, such as when choosing $D$ physically
motivated generator coordinates. The coupled cluster equations
then give a $D^6$ scaling, once intermediates 
have been calculated.

The mapping from GCM to CC provides one noteworthy
point for the coupled cluster approach,
illustrating the severity of conventional coupled cluster truncation schemes. 
In the case of CCSD, singles and doubles amplitudes $\operator{T}_{1}$ 
and $\operator{T}_{2}$ are evaluated but all higher order
amplitudes $\operator{T}_{k}$ for $k>2$ are omitted (zeroed). 
This places a strong constraint 
on the pseudo-distribution $f(\mat{Z})$ which results in 
unphysical artifacts in the resulting self-consistent CCSD equations.
One avenue for further investigation is how the 
pseudo-distributional structures for $\operator{T}_{3}$ 
and $\operator{T}_{4}$ can be exploited to address this deficiency.

\section*{Acknowledgements}
The author would like to thank Dr. Sascha Vogel for access to the Frankfurt
Institute for Advanced Studies~(FIAS) research library and 
fruitful discussions.

\appendix


\begin{thebibliography}{10}
\expandafter\ifx\csname bibnamefont\endcsname\relax
  \def\bibnamefont#1{#1}\fi
\expandafter\ifx\csname bibfnamefont\endcsname\relax
  \def\bibfnamefont#1{#1}\fi
\expandafter\ifx\csname url\endcsname\relax
  \def\url#1{\texttt{#1}}\fi
\expandafter\ifx\csname urlprefix\endcsname\relax\def\urlprefix{URL }\fi
\providecommand{\bibinfo}[2]{#2}
\providecommand{\eprint}[2][]{\url{#2}}

\bibitem{hillwheeler:1953}
\bibinfo{author}{\bibfnamefont{D.~L.} \bibnamefont{Hill}} \bibnamefont{and}
  \bibinfo{author}{\bibfnamefont{J.~A.} \bibnamefont{Wheeler}},
  \bibinfo{journal}{Phys. Rev.} \textbf{\bibinfo{volume}{89}},
  \bibinfo{pages}{1102} (\bibinfo{year}{1953}).

\bibitem{griffinwheeler:1957}
\bibinfo{author}{\bibfnamefont{J.~J.} \bibnamefont{Griffin}} \bibnamefont{and}
  \bibinfo{author}{\bibfnamefont{J.~A.} \bibnamefont{Wheeler}},
  \bibinfo{journal}{Phys. Rev.} \textbf{\bibinfo{volume}{108}},
  \bibinfo{pages}{311} (\bibinfo{year}{1957}).

\bibitem{griffin:1957}
\bibinfo{author}{\bibfnamefont{J.~J.} \bibnamefont{Griffin}},
  \bibinfo{journal}{Phys. Rev.} \textbf{\bibinfo{volume}{108}},
  \bibinfo{pages}{328} (\bibinfo{year}{1957}).

\bibitem{reinhardgoeke:1987}
\bibinfo{author}{\bibfnamefont{P.-G.} \bibnamefont{Reinhard}} \bibnamefont{and}
  \bibinfo{author}{\bibfnamefont{K.}~\bibnamefont{Goeke}},
  \bibinfo{journal}{Rep. Prog. Phys.} \textbf{\bibinfo{volume}{50}},
  \bibinfo{pages}{1} (\bibinfo{year}{1987}).

\bibitem{ringschuck:1980}
\bibinfo{author}{\bibfnamefont{P.}~\bibnamefont{Ring}} \bibnamefont{and}
  \bibinfo{author}{\bibfnamefont{P.}~\bibnamefont{Schuck}},
  \emph{\bibinfo{title}{The Nuclear Many-Body Problem}}
  (\bibinfo{publisher}{Springer-Verlag, Berlin}, \bibinfo{year}{1980}).

\bibitem{glauber:1963}
\bibinfo{author}{\bibfnamefont{R.~J.} \bibnamefont{Glauber}},
  \bibinfo{journal}{Phys. Rev.} \textbf{\bibinfo{volume}{131}},
  \bibinfo{pages}{2766} (\bibinfo{year}{1963}).

\bibitem{thouless:1960}
\bibinfo{author}{\bibfnamefont{D.~J.} \bibnamefont{Thouless}},
  \bibinfo{journal}{Nucl. Phys} \textbf{\bibinfo{volume}{21}},
  \bibinfo{pages}{225} (\bibinfo{year}{1960}).

\bibitem{thouless:1961}
\bibinfo{author}{\bibfnamefont{D.~J.} \bibnamefont{Thouless}},
  \emph{\bibinfo{title}{The quantum mechanics of many-body systems}}
  (\bibinfo{publisher}{Academic Press}, \bibinfo{year}{1961}).

\bibitem{naimark:1982}
\bibinfo{author}{\bibfnamefont{M.~A.} \bibnamefont{Naimark}} \bibnamefont{and}
  \bibinfo{author}{\bibfnamefont{A.~I.} \bibnamefont{\v{S}tern}},
  \emph{\bibinfo{title}{Theory of Group Representations}}
  (\bibinfo{publisher}{Springer-Verlag, New York, NY}, \bibinfo{year}{1982}).

\bibitem{stuber:2009}
\bibinfo{author}{\bibfnamefont{J.~L.} \bibnamefont{Stuber}},
  \bibinfo{journal}{J. Chem. Phys.} \textbf{\bibinfo{volume}{130}},
  \bibinfo{pages}{201101} (\bibinfo{year}{2009}).

\bibitem{cizek:1966}
\bibinfo{author}{\bibfnamefont{J.}~\bibnamefont{\v{C}i\v{z}ek}},
  \bibinfo{journal}{J. Chem. Phys.} \textbf{\bibinfo{volume}{45}},
  \bibinfo{pages}{4256} (\bibinfo{year}{1966}).

\bibitem{cizek:1969}
\bibinfo{author}{\bibfnamefont{J.}~\bibnamefont{\v{C}i\v{z}ek}},
  \bibinfo{journal}{Adv. Chem. Phys.} \textbf{\bibinfo{volume}{14}},
  \bibinfo{pages}{35} (\bibinfo{year}{1969}).

\bibitem{paldus:1992}
\bibinfo{author}{\bibfnamefont{J.}~\bibnamefont{Paldus}},
  \emph{\bibinfo{title}{Methods in Computational Molecular Physics, NATO ASI,
  Series B: Physics}} (\bibinfo{publisher}{Plenum Press, New York},
  \bibinfo{year}{1992}), vol. \bibinfo{volume}{293}, pp.
  \bibinfo{pages}{99--194}.

\bibitem{paldus:1999}
\bibinfo{author}{\bibfnamefont{J.}~\bibnamefont{Paldus}} \bibnamefont{and}
  \bibinfo{author}{\bibfnamefont{X.}~\bibnamefont{Li}}, \bibinfo{journal}{Adv.
  Chem. Phys.} \textbf{\bibinfo{volume}{110}}, \bibinfo{pages}{1}
  (\bibinfo{year}{1999}).

\bibitem{fukutome:1982}
\bibinfo{author}{\bibfnamefont{H.}~\bibnamefont{Fukutome}},
  \bibinfo{journal}{Prog. Thoer. Phys.} \textbf{\bibinfo{volume}{52}},
  \bibinfo{pages}{115} (\bibinfo{year}{1974}).

\bibitem{stuber:2002}
\bibinfo{author}{\bibfnamefont{J.~L.} \bibnamefont{Stuber}},
  \emph{\bibinfo{title}{Broken symmetry Hartree-Fock solutions and the
  many-electron correlation problem}}, Ph.D. thesis,
  \bibinfo{school}{University of Waterloo} (\bibinfo{year}{2002}).

\bibitem{stuber:2003}
\bibinfo{author}{\bibfnamefont{J.~L.} \bibnamefont{Stuber}} \bibnamefont{and}
  \bibinfo{author}{\bibfnamefont{J.}~\bibnamefont{Paldus}},
  \emph{\bibinfo{title}{Fundamental World of Quantum Chemistry: A Tribute
  Volume to the Memory of Per-Olov L\"{o}wdin}} (\bibinfo{publisher}{Kluwer,
  Dordrecht}, \bibinfo{year}{2003}), vol.~\bibinfo{volume}{1},
  chap.~\bibinfo{chapter}{4}, pp. \bibinfo{pages}{67--139}.

\bibitem{stuber:2015b}
\bibinfo{author}{\bibfnamefont{J.~L.} \bibnamefont{Stuber}},
  \emph{\bibinfo{title}{Explicit {E}xpressions for {P}rojected {H}artree-{F}ock via
  {T}ensor {O}perators (working title)}}, \bibinfo{note}{in preparation, expected
  on arXiv before 2015-07-01}.

\end{thebibliography}

\end{document}